# Fine-tuned Generative Adversarial Network-based Model for Medical Image Super-Resolution

Alireza Aghelan, Modjtaba Rouhani
Computer Engineering Department, Ferdowsi University of Mashhad
Mashhad, Iran
aghelan.alireza@mail.um.ac.ir, rouhani@um.ac.ir

***Abstract*— In the field of medical image analysis, there is a substantial need for high-resolution (HR) images to improve diagnostic accuracy. However, it is a challenging task to obtain HR medical images, as it requires advanced instruments and significant time. Deep learning-based super-resolution methods can help to improve the resolution and perceptual quality of low-resolution (LR) medical images. Recently, Generative Adversarial Network (GAN) based methods have shown remarkable performance among deep learning-based super-resolution methods. Real-Enhanced Super-Resolution Generative Adversarial Network (Real-ESRGAN) is a practical model for recovering HR images from real-world LR images. In our proposed approach, we use transfer learning technique and fine-tune the pre-trained Real-ESRGAN model using medical image datasets. This technique helps in improving the performance of the model. We employ the high-order degradation model of the Real-ESRGAN which better simulates real-world image degradations. This adaptation allows for generating more realistic degraded medical images, resulting in improved performance. The focus of this paper is on enhancing the resolution and perceptual quality of chest X-ray and retinal images. We use the Tuberculosis chest X-ray (Shenzhen) dataset and the STARE dataset of retinal images for fine-tuning the model. The proposed model achieves superior perceptual quality compared to the Real-ESRGAN model, effectively preserving fine details and generating images with more realistic textures.**

## I. Introduction

Medical images are used to diagnose diseases and analyze and retrieve important information. In medical image analysis, a significant requirement exists for abundant details in the images. Usually, a better diagnosis is achieved if the image has a high resolution and fine details are preserved in the image. Unfortunately, capturing high-resolution (HR) medical images is costly and challenging, as it needs complex and expensive equipment, skilled personnel, and a considerable amount of time. [1] To address these issues, we can use accurate super-resolution models as a post-processing step to recover HR images from low-resolution (LR) medical images. [2]

Preserving fine details and removing artifacts is an important issue in the field of medical image super-resolution that requires special attention. Fine details in medical images are of significant importance, as they contain valuable information for diagnosing diseases. For example, fine vessels in retinal images are used to detect the swelling of vessels. Furthermore, in brain MRI images, fine details around a tumor help to identify the origin of the tumor. Therefore, removing important details and producing artifacts by medical image super-resolution models may unfavorably affect the performance of medical image interpretation and lead to misdiagnosis. [1]

Traditional super-resolution methods are divided into three categories: interpolation-based methods, reconstruction-based methods, and shallow learning-based methods. [3] Interpolation-based methods frequently cannot recover high-frequency content and produce blurred edges. [4] Reconstruction-based and shallow learning-based methods have better performance than interpolation-based methods. However, these methods have some limitations in reconstructing high-frequency information.

Deep learning-based super-resolution methods have superior performance compared to most traditional methods. [1] These methods can effectively recover high-frequency details of LR images, resulting in a considerable improvement in their perceptual quality. At first, they only used Convolutional Neural Networks (CNNs). CNN-based super-resolution methods often produce HR outputs that lack clear edges and rich texture details. Nowadays, Generative Adversarial Network (GAN) [5] based methods are widely used in general and medical image super-resolution fields. These methods have shown significant performance in generating realistic outputs close to the ground truth HR image.

General GAN-based super-resolution models achieve good HR results for natural images; however, they often fail to produce satisfactory results when applied to medical images. In fact, due to the different distributions of medical and natural images, a model trained on natural image datasets frequently exhibits suboptimal performance for medical image super-resolution. To address this issue, we can train GAN-based super-resolution models from scratch with medical image datasets. However, this approach is not efficient enough, as most medical datasets contain a limited number of images that do not have a very high spatial resolution. A more effective solution is to use the transfer learning technique and fine-tune pre-trained models on medical image datasets. This technique is useful in medical applications and can improve the performance of the model. [4]

Real-Enhanced Super-Resolution Generative Adversarial Network (Real-ESRGAN) [6] is an efficient model for enhancing the resolution and perceptual quality of real-world images. The remarkable performance of the Real-ESRGAN model and the advantages of transfer learning, motivated us to fine-tune this model for medical image super-resolution. This model utilizes a high-order degradation model to better simulate real-world degradations in natural images. Although this degradation model was developed for natural images, it can accurately emulate noise, blur, and artifacts in medical images. Therefore, we use this degradation model to improve the performance of our model. In this paper, we aim to improve the resolution and perceptual quality of chest X-ray and retinal images. In our proposed approach, the pre-trained Real-ESRGAN model is fine-tuned on chest X-ray and retinal datasets, separately. The fine-tuned model can recover more realistic textures and better preserve fine details compared to the Real-ESRGAN model.

The rest of this paper is organized as follows. We go through some related literature in Section 2. Section 3 includes the datasets used for fine-tuning, the architecture of the Real-ESRGAN model, a review of the degradation process for generating degraded LR images, the benefits of using the transfer learning technique, and fine-tuning details. Section 4 is devoted to the review and analysis of the results, and Conclusion remarks and future directions are given in Section 5.

## II. Related Work

Due to the different distributions of medical and natural images, separate super-resolution models have been developed for each of them. In this section, several related medical and general GAN-based super-resolution models are reviewed separately.

### A. *Medical image super-resolution models*

In some recently proposed GAN-based approaches for medical image super-resolution, researchers have used transfer learning technique, and their proposed models have shown excellent performance. In this section, we review some GAN-based medical image super-resolution models that utilized transfer learning.

In a related study focusing on enhancing the resolution and visual quality of periapical radiographs, Morana et al. [7] employed transfer learning and the super-resolution generative adversarial network (SRGAN) [8] model. The authors performed separate pre-training processes on a chest X-ray dataset and the "102-flowers" dataset. Their findings showed that this approach can compensate for the shortage of medical training data and improve model convergence and performance. Gupta et al. [2] also used transfer learning in their GAN-based MRI image super-resolution model, resulting in improved model performance. In Lv et al. [9] the authors investigated the application of transfer learning to enhance GAN-based multi-channel MRI reconstruction despite limited medical training data. They pre-trained their model on public Calgary brain images and fine-tuned it on various medical image datasets, including brain tumor patients and different anatomies like knee and liver. Their results demonstrated transfer learning's ability to improve reconstruction quality and achieve effective generalization to unseen data. For multimodal medical image super-resolution, Dharejo et al. [4] proposed a multi-attention GAN framework with wavelet transform. The authors first trained their model on DIV2K [10] dataset and then fine-tuned it on medical image datasets. They reported similar improvements that underscore the efficacy of transfer learning. Sun et al. [11] fine-tuned the pre-trained Real-ESRGAN model to reduce blooming artifacts in calcified coronary plaques. Their fine-tuned model significantly improves the diagnostic performance of coronary computed tomography angiography (CCTA) in plaque assessment.

The reviewed models clearly demonstrate the benefits of applying transfer learning for medical image super-resolution. More details on the specific advantages of using transfer learning can be found in Section 3.4. Inspired by these findings, we utilize this technique in our study.

### B. *General image super-resolution models*

In this section, we provide an overview of SRGAN [8] and ESRGAN [12] models and their advancements in improving the perceptual quality of real-world images. We also discuss the Real-ESRGAN model and its ability to simulate real-world degradations.

SRGAN, proposed by Ledig et al. [8], is a GAN-based model that showed substantial improvement in the field of single image super-resolution (SISR). The SRGAN generator is a deep residual network (ResNet) with skip connections. Utilization of an enhanced perceptual loss function that is composed of adversarial loss and content loss significantly improved the performance of this model. To design a content loss that more accurately measures the perceptual similarity, the authors used feature maps of the VGG network. The employment of adversarial loss helped to produce more realistic and natural images. The SRGAN model generates images with better perceptual quality compared to previous models. However, the outputs of this model often contain unpleasant artifacts. To further improve the perceptual quality of the SRGAN outputs, Wang et al. [12] applied several improvements to this model and presented the ESRGAN model. The authors proposed the Residual-in-Residual Dense Block (RRDB) based generator without batch normalization layers. The RRDB structure has more depth and intricacy compared to the residual block structure of the SRGAN model. Removing the batch normalization layers resulted in improved performance and reduced artifacts. They employed residual scaling and smaller initialization to make network training easier. In the ERSGAN model, the standard SRGAN discriminator was replaced by a relativistic discriminator [13]. They also created a better perceptual loss by using the features prior to activation. These improvements made the ESRGAN model perform better in restoring texture details and removing unpleasant artifacts.

In the field of blind image super-resolution, much research has been conducted on different degradation models. Researchers have tried to recover HR images from LR images that are degraded by complex and unknown degradations.

However, their degradation models usually cannot simulate complex real-world degradations. To address this issue, Wang et al. [6] proposed the Real-ESRGAN model that uses a high-order degradation model to simulate more practical degradations. One of the advantages of the high-order degradation model is the ability to simulate ringing and overshoot artifacts. The ringing and overshoot artifacts are usually generated by JPEG compression, etc. More details of the high-order degradation model are described in Section 3.3. In the Real-ESRGAN model, a U-Net discriminator [14] that incorporates spectral normalization [15] is utilized to enhance discrimination power and stabilize the training process. Benefiting from these improvements, the Real-ESRGAN model achieves superior perceptual quality than previous models and can be used in real-world applications.

The Real-ESRGAN model has been successfully applied in various domains, including underwater image super-resolution [16]. However, its application in medical image super-resolution domain remains relatively unexplored. In this paper, we extend the capabilities of the Real-ESRGAN model to medical image super-resolution, enhancing the resolution and perceptual quality of chest X-ray and retinal images.

## III. Methodology

This section consists of the following sub-sections: The first sub-section includes the important points in dataset selection and the datasets used for fine-tuning. The architecture of the generator and discriminator networks of the Real-ESRGAN model and its improvements compared to previous models are reviewed in sub-section 2. In the next sub-section, the method used for generating degraded LR images is discussed. Sub-section 4 explains the advantages of using transfer learning technique and the process of fine-tuning the model. The details of the fine-tuning process are described in the last sub-section.

### *A. Datasets*

One of the important steps in fine-tuning is choosing the appropriate dataset. We have considered two points in selecting the appropriate dataset for fine-tuning GAN-based super-resolution models: 1) the selected dataset should contain high-quality and high-resolution images; 2) in most GAN-based methods, a substantial amount of data makes fine-tuning the model more effective. Therefore, it is advantageous that the selected dataset contains a relatively large amount of data.

Retinal images dataset. STARE (STructured Analysis of the Retina) dataset is one of the most widely used datasets in medical applications. This dataset contains 397 images. Each image in this dataset has a resolution of 700 × 605. We use the original images as HR images. Degraded LR images are generated by the high-order degradation process of the Real-ESRGAN model. The degradation process is explained in Section 3.3.

Chest X-ray images dataset. Tuberculosis chest X-ray (Shenzhen) dataset contains high-resolution images that are suitable for our work. This dataset includes 662 images with different resolutions. The original images are used as ground truth images. We produce multi-scale images from the original images. The detailed process of generating multi-scale images is explained in the next paragraph. To produce degraded LR images, we use the high-order degradation process of the Real-ESRGAN model.

Most of the images in the chest X-ray dataset have a resolution ranging from approximately 2.5K × 2.5K to 3K × 3K pixels. For this dataset, we produce multi-scale images from the original images to fine-tune the model with images of varying target resolutions. Generating multi-scale images results in improved adaptability and performance across different scales. This process involves downsampling the original images using the Lanczos method to produce several ground truth images at different scales. These multi-scale images are then combined with the original images. Ultimately, the number of ground truth images increases fivefold.

To demonstrate the diversity and complexity of the datasets used to fine-tune the model, we provide an overview of sample images from these datasets in Fig. 1. We present a selection of representative images from the Tuberculosis chest X-ray dataset and sample retinal images from the STARE dataset. Furthermore, Fig. 1 exhibits examples of the natural images that were used by Wang et al. [6] to train the Real-ESRGAN model from scratch. The natural images in Fig. 1 have been selected from DIV2K [10] dataset. The datasets were utilized for training the model are mentioned in Section 3.4.

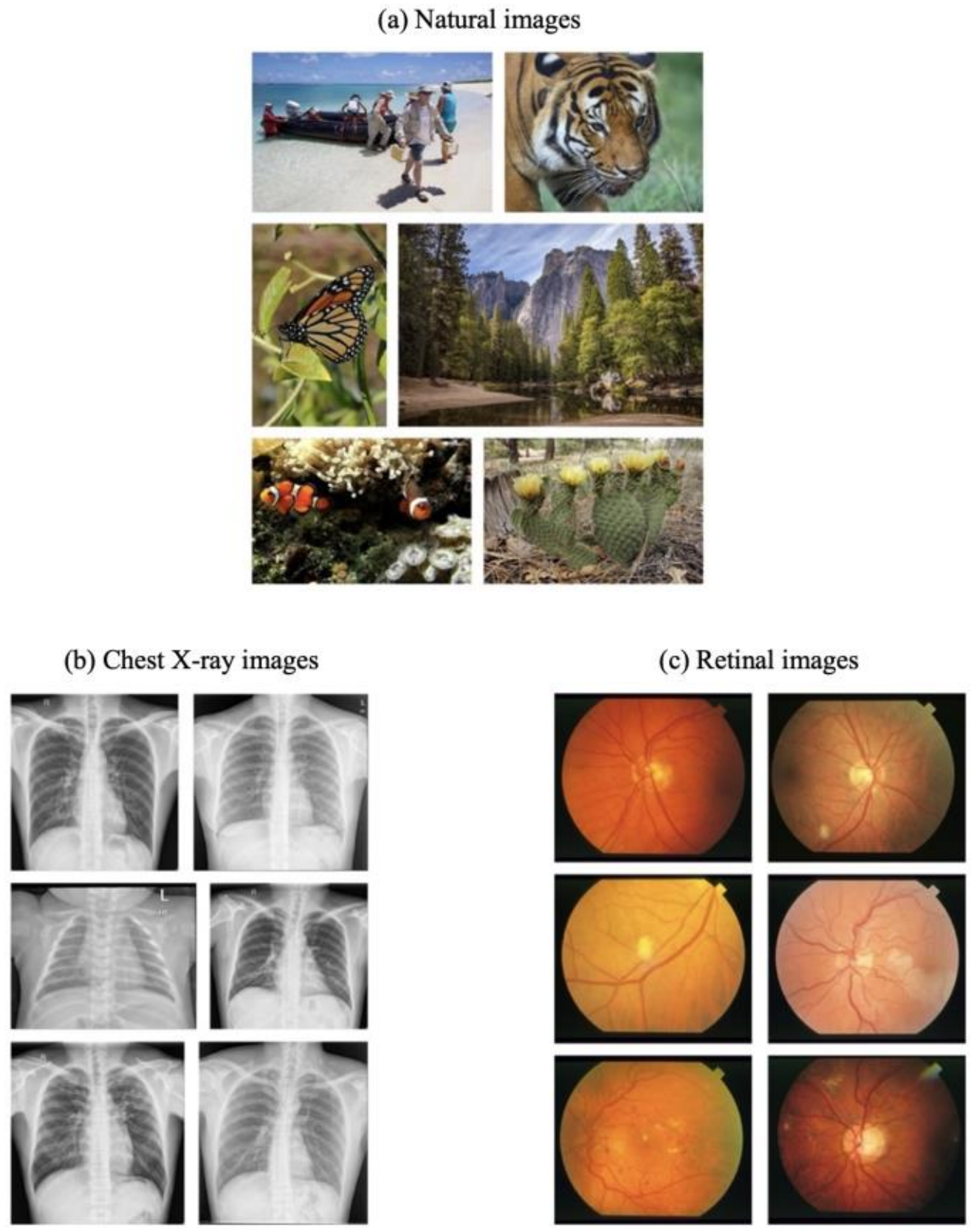

Fig. 1. Sample images from datasets. (a) Natural images were utilized to train the model. (b) Chest X-ray images are used for fine-tuning the model. (c) Retinal images are employed for fine-tuning the model.

## B. Networks Architecture

In Wang et al. [6] the authors employed the RRDB-based generator of the ESRGAN [12] model. The architecture of this generator is shown in Fig. 2. The employed generator has 23 RRDB blocks, and each of these RRDB blocks consists of three dense blocks. More details about the RRDB-based generator can be found in the ESRGAN paper [12]. The original ESRGAN model was designed for 4× upscaling. To expand upon this limitation, the Real-ESRGAN model offers the capability to perform super-resolution with various scaling factors such as 1×, 2×, and 4×.

One of the improvements of the Real-ESRGAN model is the use of a U-Net discriminator [14] with skip connections. The U-Net based discriminator can provide global and local (per-pixel) feedback to the generator. This discriminator has more discriminative power compared to the VGG-style discriminator in the ESRGAN model. The utilized discriminator helps in enhancing local details. In the Real-ESRGAN model, spectral normalization [15] is employed to stabilize the training of the discriminator. This technique is also useful for reducing overly sharp and unpleasant artifacts.

## C. Degradation model

We can generate degraded LR images from ground truth HR images in two ways: 1- Employing the high-order degradation process of the Real-ESRGAN model. 2- Using traditional degradation methods. Traditional degradation process usually includes blurring, applying various noises, downsampling, and JPEG compression. To better simulate real-world degradations, Wang et al. [6] developed the traditional degradation model into a high-order degradation process. As mentioned before, ringing and overshoot artifacts have been considered in the high-order degradation process. To synthesize these artifacts, the authors used a 2D sinc filter that cuts off high frequencies. The high-order degradation process helps to recover realistic textures and remove artifacts, resulting in a remarkable improvement in the model's performance. Therefore, we use the first method to generate degraded LR images. In the first method, degraded LR images are generated by the high-order degradation process during fine-tuning. Visual representation of the high-order degradation process is shown in Fig. 3.

In medical imaging, various factors contribute to the degradation of image quality, such as restricted spatial resolution, random noise, motion-induced blurring, and imaging artifacts like beam hardening and partial volume effects. The high-order degradation process is particularly beneficial in this context, as it simulates complex real-world degradations more accurately than traditional methods. Using the high-order degradation process helps the model preserve critical anatomical details like fine vessels in retinal images or subtle abnormalities in chest X-rays while efficiently reducing noise and artifacts. This leads to improved image quality, robustness, and generalization, making the high-order degradation process essential for medical image super-resolution.

## D. Transfer learning

Training the Real-ESRGAN model from scratch requires powerful hardware resources, a huge number of images, and a significant amount of time. In fact, the Real-ESRGAN model has been trained with four NVIDIA V100 GPUs on DIV2K [10], Flickr2K [17], and OutdoorSceneTraining [18] datasets. One of the significant challenges in the field of medical image super-resolution is the relatively small amount of data in most medical datasets. In addition, we need high-quality and high-resolution images for training super-resolution models. However, most medical datasets do not have very high spatial resolution images. To overcome these challenges, we employ transfer learning technique.

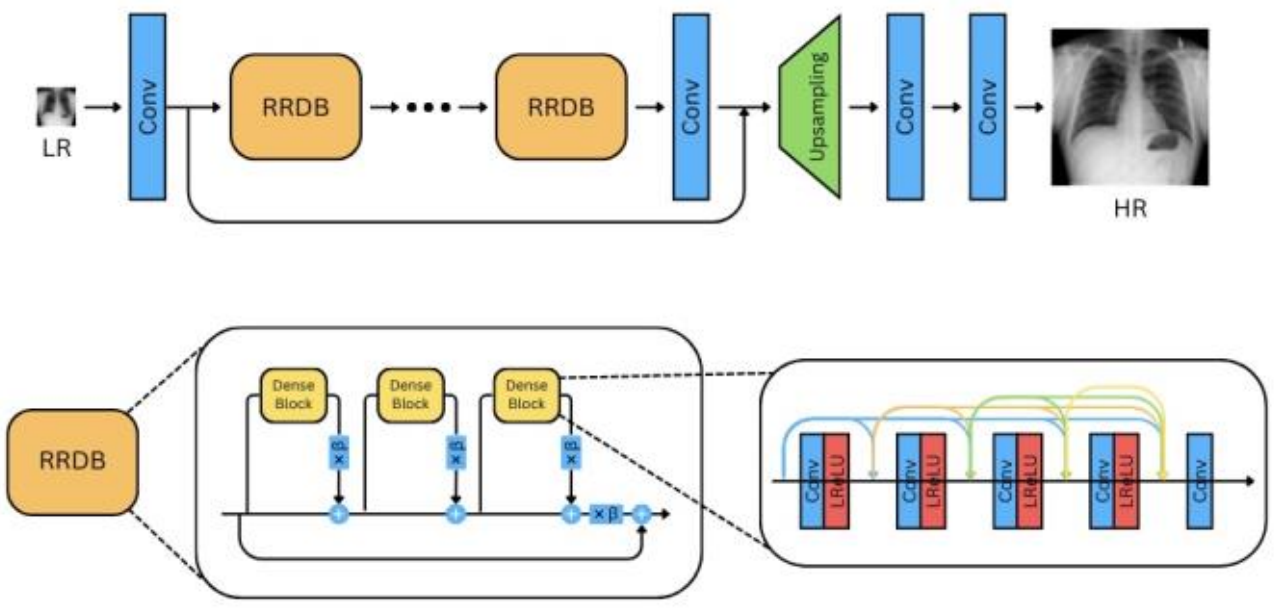

Fig. 2. Generator architecture of the ESRGAN model.

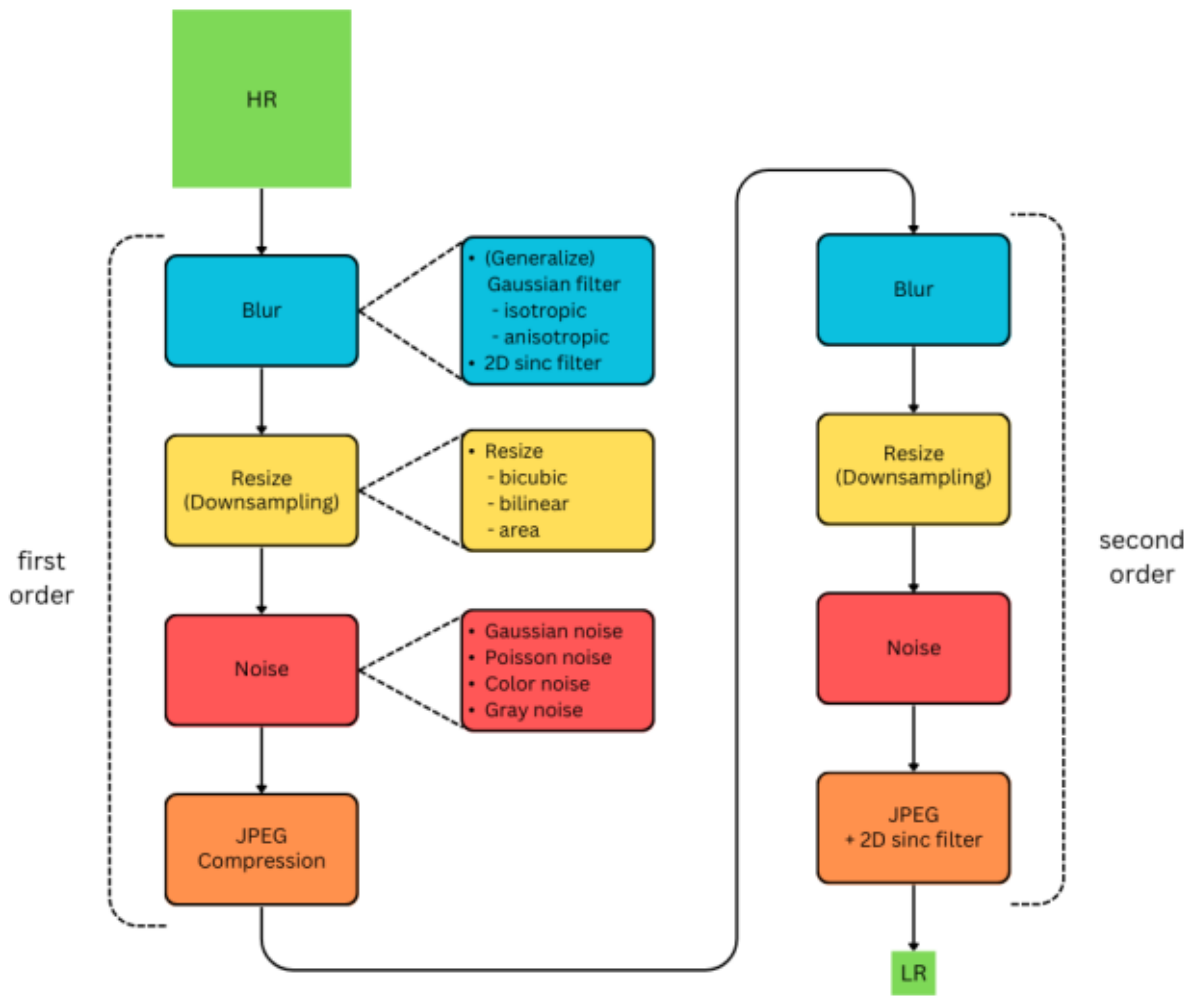

Fig. 3. Diagram illustrating the high-order degradation process of the Real-ESRGAN model.

In this work, transfer learning is the process of taking the Real-ESRGAN model that has been trained for general image super-resolution, and then fine-tuning it for medical image super-resolution on medical image datasets. Transfer learning is useful in medical applications and can compensate for the shortage of data. By using knowledge gained from natural image datasets, this technique can improve the performance of the model. [4] Furthermore, transfer learning helps the model to converge faster by providing a strong initialization based on the pre-trained weights. Additionally, this technique can improve the model's generalization capability, allowing it to effectively perform on previously unseen data.

In our proposed approach, the pre-trained generator (v0.1.0) and discriminator (v0.2.2.3) networks of the RealESRGAN_x4plus model are first loaded from GitHub. In the next step, the loaded networks are fine-tuned using medical image datasets. Due to the different distributions of chest X-ray and retinal images, we fine-tune the pre-trained Real-ESRGAN model separately on chest X-ray and retinal datasets and provide two super-resolution models for chest X-ray and retinal images. Fig. 4 shows the key steps involved in the fine-tuning process of the model.

Due to the Real-ESRGAN model has been trained on datasets consisting of natural images, which have different distributions compared to medical images, a substantial number of epochs is necessary to effectively fine-tune the model. This extended fine-tuning time enables the model to better accommodate and optimize the specific characteristics and nuances of medical images, resulting in improved performance in generating more accurate and realistic outputs. The number of epochs used for fine-tuning the model on chest X-ray and retinal datasets can be found in Section 3.5.

### E. *Fine-tuning details*

In the Real-ESRGAN model, the generator network has 16,697,987 parameters and the discriminator network has 4,376,897 parameters. We use Adam optimizer with a learning rate of 0.0001 to optimize the weights of these networks. In this process, a combination of L1 loss, perceptual loss, and GAN loss functions is utilized. In addition, exponential moving average (EMA) is employed for more stable fine-tuning. We also adopt the same strategy as Real-ESRGAN+ by sharpening the ground truth images with unsharp masking (USM) during fine-tuning, which helps achieve an optimal balance between enhanced sharpness and suppression of overshoot artifacts.

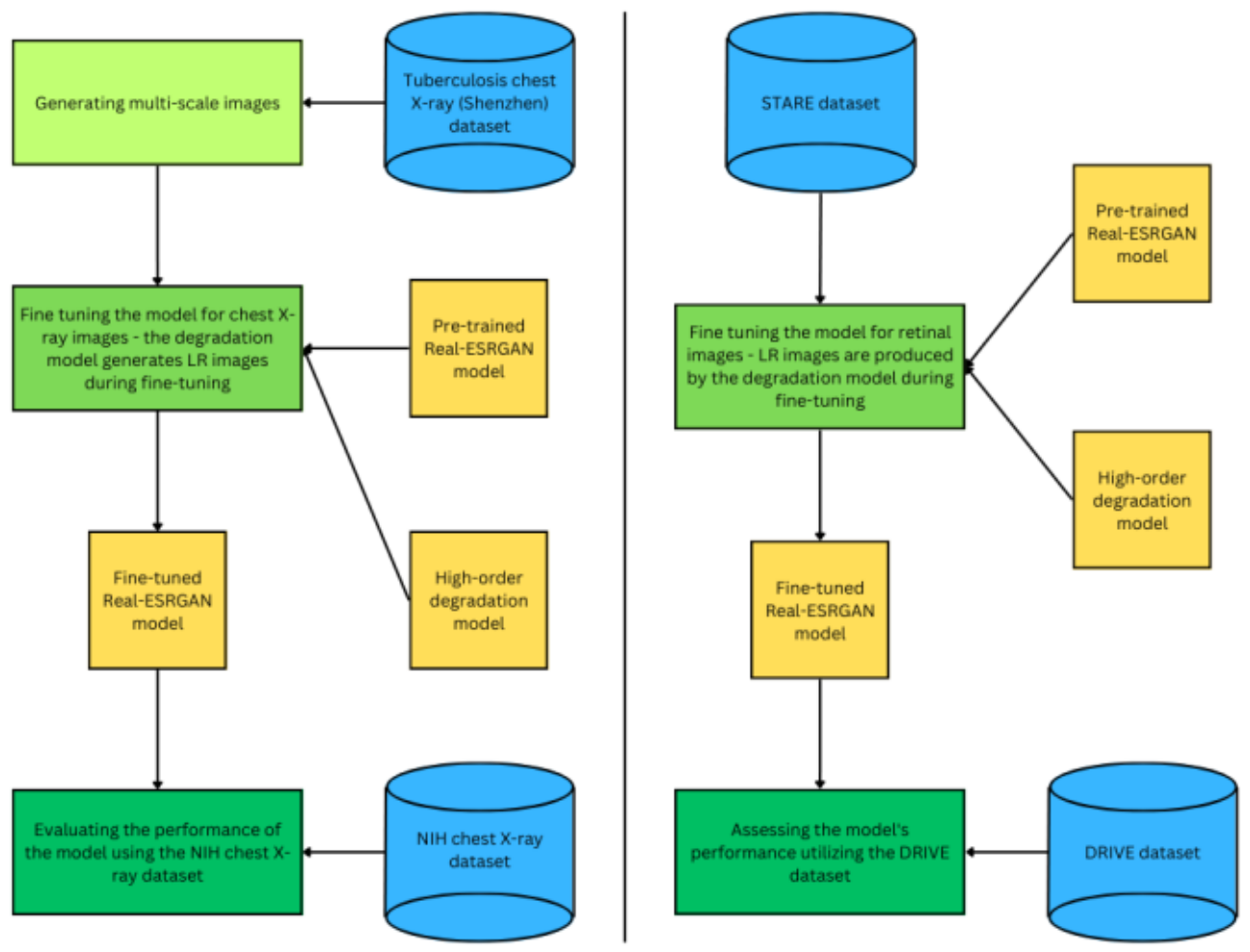

Fig. 4. Flowchart demonstrating the fine-tuning process of the pre-trained Real-ESRGAN model for medical image super-resolution. Left side: Fine-tuning on the chest X-ray dataset. Right side: Fine-tuning on the retinal dataset.

For the retinal dataset, we fine-tune the pre-trained Real-ESRGAN model for 3000 iterations, which are approximately equal to 75 epochs. Furthermore, we fine-tune the pre-trained Real-ESRGAN model on the chest X-ray dataset for 16600 iterations (approximately 50 epochs). In this work, Google Colab GPU with a batch size of 10 is used to fine-tune the model.

## IV. Results

Most previous models for medical image super-resolution have mainly focused on MRI and CT images and it is not feasible to compare them with our model. In this section, we provide a comprehensive comparison between our fine-tuned model and the baseline Real-ESRGAN model to demonstrate the improvements resulting from optimizing the model specifically for chest X-ray and retinal datasets. Furthermore, we conduct an adequate analysis of the results and discuss the clinical implications of the findings.

To ensure a comprehensive assessment of the models, we employ both quantitative and qualitative evaluation methods. However, quantitative metrics, such as peak signal-to-noise ratio (PSNR) and Structural Similarity Index (SSIM), cannot accurately reflect the perceptual preferences of humans. These metrics are insufficient for evaluating fine details, such as detailed texture information. In fact, achieving high PSNR and SSIM values does not necessarily ensure superior perceptual quality. For example, in Fig. 7, the PSNR and SSIM values for images in column 3 are as follows: the Real-ESRGAN output has a PSNR of 29.49 and SSIM of 0.7928, while the fine-tuned model output has a PSNR of 32.41 and SSIM of 0.7699. As one can see, despite the higher SSIM value for the Real-ESRGAN model, its output contains unrealistic details. On the other hand, the fine-tuned model generates more realistic outputs that are closer to the ground truth image. Considering these limitations, we place significant emphasis on qualitative results for a more accurate evaluation of the models. Qualitative evaluation allows us to assess the preservation of fine details, as well as the realism and naturalness of the generated images. By considering these factors, we obtain a more comprehensive understanding of the models' performance.

The performance evaluation process and results analysis for the chest X-ray and retinal images are presented in separate sub-sections below.

### A. *Retinal images*

To evaluate the performance of the model, we use the DRIVE dataset. This dataset contains 40 retinal images with a resolution of 565 × 584. We rescale each of these images to 512 × 512 resolution using the Lanczos method, which effectively preserves image quality and is ideal for generating ground truth images. These rescaled images are used as ground truth images. Degraded LR images are obtained from the ground truth images using a two-step process. First, the images are downsampled by a factor of 2 using the bicubic method, which is widely used in super-resolution studies for standardized comparisons. Then, Gaussian blurring is applied to the downsampled images. To enhance the resolution of the degraded LR images, we use an upsampling scale factor of 2. Quantitative results are demonstrated in Table 1, and qualitative results for three different retinal images are shown in Fig 5 & 6.

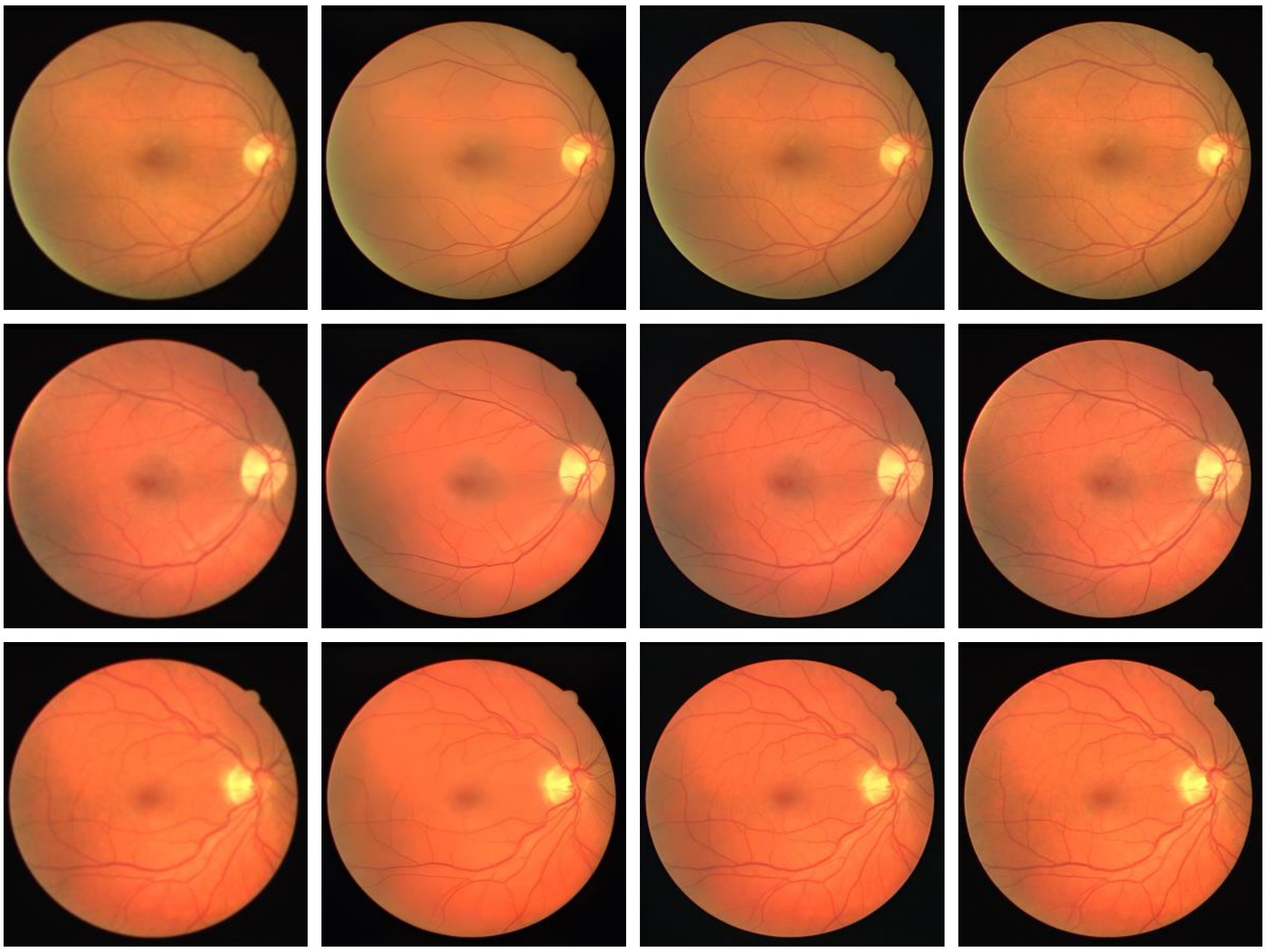

Fig. 5. Qualitative results for images of the DRIVE dataset. Columns 1–4 are degraded LR images, Real-ESRGAN outputs, fine-tuned Real-ESRGAN outputs, and ground truth HR images.

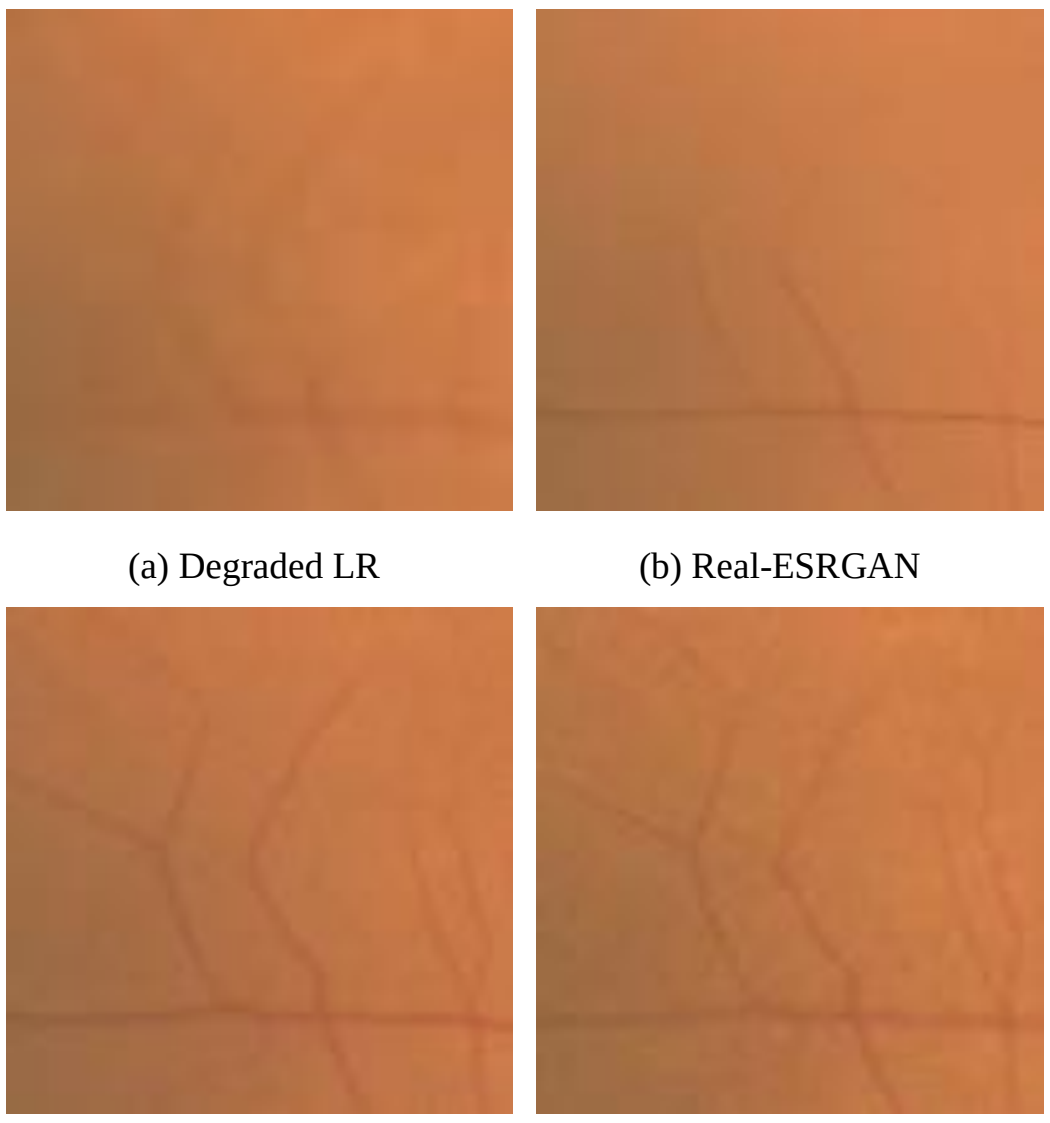

(a) Degraded LR (b) Real-ESRGAN

(c) Fine-tuned Real-ESRGAN (d) Ground Truth

Fig. 6. Magnified regions of a retinal image. (a) Degraded LR image. (b) Real-ESRGAN output. (c) Fine-tuned Real-ESRGAN output. (d) Ground truth HR image.

| **Model** | **Quantitative Results** | |
|---|---|---|
| | ***PSNR*** | ***SSIM*** |
| Real-ESRGAN | **35.33** | 0.8866 |
| Fine-tuned Real-ESRGAN | 34.21 | **0.8923** |

TABLE 1

In quantitative results, the Real-ESRGAN model exhibits higher PSNR values compared to the fine-tuned model, while the fine-tuned model shows superior SSIM values. However, these metrics alone do not provide a conclusive determination of the superiority between the models. This emphasizes the significance of qualitative evaluation in assessing the perceptual quality of the generated images.

As shown in Fig. 5, the Real-ESRGAN model cannot effectively preserve fine details, including fine vessels. In comparison, the fine-tuned model shows exceptional accuracy in preserving fine details and recovers more realistic textures.

Fine details, notably fine vessels, are of remarkable importance in retinal image analysis. Fine vessels serve as crucial indicators for diagnosing retinal conditions such as diabetic retinopathy, age-related macular degeneration, and retinal vascular occlusions. By preserving delicate structures like fine vessels, the identification of subtle changes in vessel morphology is improved. This improvement enables more accurate analysis and diagnosis and leads to improved clinical decision-making.

### *B. Chest X-ray images*

We utilize the NIH chest X-ray dataset to test the models. The resolution of each image in this dataset is 1024 × 1024. We randomly select 40 images from this dataset and use them as ground truth images. In order to generate degraded LR images from the ground truth images, a two-step process involving 4× bicubic downsampling followed by Gaussian blurring is employed. We increase the resolution of the degraded LR images with a scale factor of 4×. Table 2 shows the quantitative results, and Fig. 7 & 8 demonstrate the qualitative results for three different chest X-ray images.

As shown in Table 2, the fine-tuned model achieves higher PSNR values, whereas the Real-ERSGAN model demonstrates superior SSIM values. Similar to the quantitative results for retinal images, these metrics cannot definitively determine model superiority, and we rely on qualitative results for a more meaningful evaluation.

It can be observed from Fig. 7 that the Real-ESRGAN outputs suffer from the inclusion of unrealistic details. In contrast, the outputs generated by the fine-tuned model exhibit a significant improvement in terms of realism. They are closer to the ground truth HR images, which shows the efficacy of the fine-tuning process in enhancing accuracy and producing images with better perceptual quality.

| **Model** | **Quantitative Results** | |
|---|---|---|
| | ***PSNR*** | ***SSIM*** |
| Real-ESRGAN | 32.09 | **0.8668** |
| Fine-tuned Real-ESRGAN | **32.46** | 0.8232 |

TABLE 1I

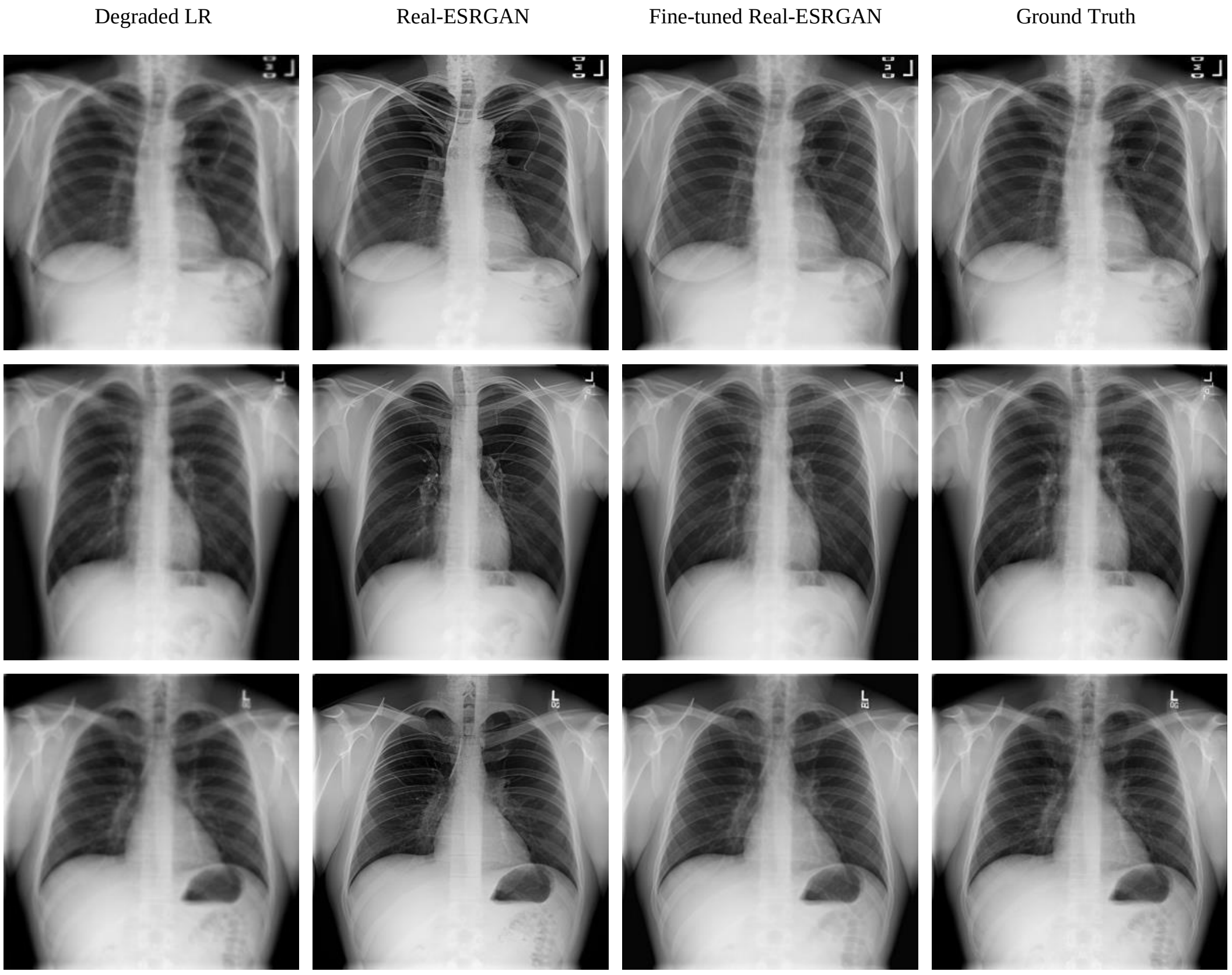

Fig. 7. Qualitative results for images of the NIH chest X-ray dataset. Columns 1–4 are degraded LR images, Real-ESRGAN outputs, fine-tuned Real-ESRGAN outputs, and ground truth HR images.

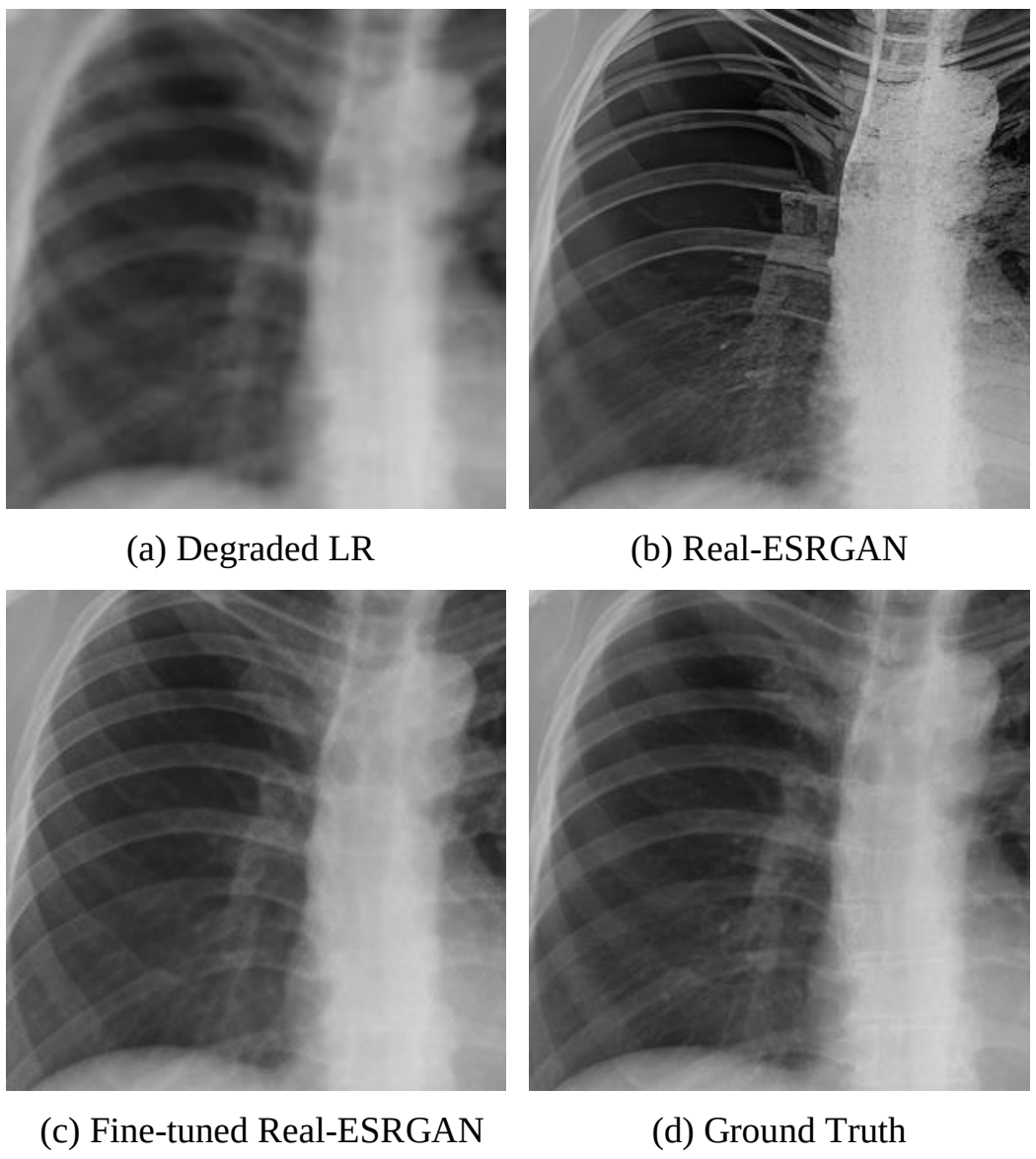

Fig. 8. Magnified regions of a chest X-ray image. (a) Degraded LR image. (b) Real-ESRGAN output. (c) Fine-tuned Real-ESRGAN output. (d) Ground truth HR image.

Enhancing the resolution and perceptual quality of chest X-ray images contributes to improving diagnostic accuracy in the interpretation of various illnesses, including pneumonia, tuberculosis, lung cancer, and others. The results demonstrate that the Real-ESRGAN model is unable to effectively reach this goal. On the other hand, the fine-tuned model achieves superior perceptual quality and enables clinicians to accurately identify subtle abnormalities, recognize indicators of diseases, and make dependable assessments.

## V. Conclusion and future directions

In this work, we modified the Real-ESRGAN model for medical image super-resolution by fine-tuning it on medical image datasets. Our fine-tuned model demonstrated remarkable improvements compared to the Real-ESRGAN model, producing outputs that are closer to the ground truth images and exhibit more realistic textures. These findings highlight the potential effectiveness of our model for real-world applications and provide valuable prospects to improve diagnostic accuracy and facilitate informed clinical decision-making.

In this paper, we worked on chest X-ray and retinal images. The Real-ESRGAN model can also be fine-tuned to improve the resolution and perceptual quality of MRI and CT scan images. One future direction is to use the A-ESRGAN [19] model and fine-tune it for medical image super-resolution. The A-ESRGAN is an improved version of the Real-ESRGAN that employs an attention U-Net based, multi-scale discriminator.